\RequirePackage{fix-cm}
\documentclass[allclo, smallextended, square, comma, sort&compress, natbib]{svjour3}
\bibpunct{[}{]}{;}{n}{}{,} 
\smartqed  
\usepackage{graphicx}
\usepackage{amsmath,amsfonts,amssymb,bm}
\usepackage{dsfont}
\usepackage{graphicx}
\usepackage{color}
\usepackage{soul}

\usepackage[mathscr,scaled=1.15]{urwchancal}
\DeclareFontFamily{OT1}{pzc}{}
\DeclareFontShape{OT1}{pzc}{m}{it}%
{<-> s * [1.15] pzcmi7t}{}
\DeclareMathAlphabet{\mathpzc}{OT1}{pzc}{m}{it}
\begin{document}

\title{Process-independent effective coupling
}
\subtitle{From QCD Green's functions to phenomenology}


\author{Jose Rodr\'{\i}guez-Quintero  \and \\ Daniele Binosi \and C\'edric Mezrag \and Joannis Papavassiliou \and Craig D. Roberts
}


\institute{J. Rodr\'{\i}guez-Quintero \at
              Department of Integrated Sciences;
University of Huelva, E-21071 Huelva, Spain \\
              \email{jose.rodriguez@dfaie.uhu.es}           
           \and
           D. Binosi \at
              European Centre for Theoretical Studies in Nuclear Physics
and Related Areas (ECT$^\ast$) and Fondazione Bruno Kessler\\ Villa Tambosi, Strada delle Tabarelle 286, I-38123 Villazzano (TN), Italy\\
\email{binosi@ectstar.eu}
			\and
			C. Mezrag \at
			Istituto Nazionale di Fisica Nucleare, Sezione di Roma, P. le A. Moro 2, I-00185 Roma, Italy\\ 
			\email{cedric.mezrag@roma1.infn.it}
			\and
			J. Papavassiliou \at 
			Department of Theoretical Physics and IFIC, University of Valencia and CSIC, E-46100, Valencia, Spain \\
			\email{Joannis.Papavassiliou@uv.es}
			\and
			C. D. Roberts \at
			Physics Division, Argonne National Laboratory, Argonne IL 60439, USA \\ 
			\email{cdroberts@anl.gov}
}

\date{Received: date / Accepted: date}

\maketitle

\begin{abstract}
This article reports on a very recent proposal for a new type of process-independent QCD effective charge~\cite{Binosi:2016nme} defined, as an anologue of the Gell-Mann-Low effective charge in QCD, on the ground of nothing but the knowledge of the gauge-field two-point Green's function, albeit modified within a particular computational framework; namely, the combination of pinch technique and background field method which makes possible a systematic rearranging of classes of diagrams in order to redefine the Green's function and have them obey linear QED-like Slavnov-Taylor identities. We have here calculated that effective charge, shown how strikingly well it compares to a process-dependent effective charge based on the Bjorken sum rule; and, finally, employed it in an exploratory calculation of the proton electromagnetic form factor in the hard scattering regime. 
 
\keywords{Confinement \and Strong interaction coupling and masses \and QCD Green's functions \and nucleon form factors}
\end{abstract}

\section{Introduction}
\label{intro}

Strong interactions in Nature are successfully described through a quantum non-Abelian gauge field theory endowed with a very rich non-perturbative sector where crucial phenomena such as confinement and dynamical chiral symmetry breaking take place; namely, quantum chromodynamics (QCD). As it also happens for its archetypical precursor as a quantum gauge field theory accounting for interactions in Nature, quantum electrodynamics (QED), its Lagrangian couplings and masses do not remain constant. They come, instead, to depend on a momentum or mass scale. Indeed, this is a general and collateral upshot of quantisation and ultraviolet renormalisation in any four-dimension quantum gauge field theory. 

In the case of QED, grounded on an Abelian gauge symmetry, the theory can be genuinely approached as a perturbation theory within which, owing to the Ward identity~\cite{Ward:1950xp}, there is a running coupling, measuring the strength of the photon--charged-fermion vertex, which can be obtained by computing the photon vacuum polarisation collecting all the virtual processes that change the bare photon into a dressed object.  

In the case of QCD~\cite{Marciano:1979wa}, the scheme might {\it a priori} be seen as analogous: the classical Lagrangian introduces four possibly distinct strong-interaction vertices for the quantised and renormalised theory, but the non-Abelian gauge invariance translates into the BRST symmetry \cite{Becchi:1975nq,Tyutin:1975qk}, {\it via} the Slavnov-Taylor identities (STIs)~\cite{Taylor:1971ff,Slavnov:1972fg}, which leaves us with a single running coupling characterising all four interactions. Though, the latter cannot be ensured on the domain within which perturbation theory is not reliable. Here, precisely, it lies the main difference of QCD regarding to QED: owing to its non-Abelian nature, the former is an asymptotically free theory for which perturbation theory is only valid at large momentum scales; while all dynamics taking place at momenta of the order of the proton's mass or below ({\it i.e.}, the typical ones for strong-interaction phenomena) is genuinely nonperturbative. Therefore, on theoretical grounds, nothing obvious prevents the existence of four distinct non-perturbative couplings defined within the IR domain which, anyhow, must all become equivalent at asymptotically large momenta on the perturbative domain. This implies also the possibility of different non-connected renormalisation-group-invariant (RGI) intrinsic mass-scales for each coupling which, in their turn, might be a source for BRST symmetry breaking and have an impact on the renormalisability of the theory. Our view is nevertheless that this is not the case and there exist a unique running coupling and an intrinsic mass-scale which, owing to the dynamical generation of gluon \cite{Cucchieri:2007md,Cucchieri:2007rg,Aguilar:2008xm,Boucaud:2008ky,Dudal:2008sp,Bogolubsky:2009dc,Aguilar:2012rz} and quark masses \cite{Bhagwat:2003vw,Bowman:2005vx,Bhagwat:2006tu}, all comparatively large at IR momenta, make sure that QCD is a well-defined theory at all momentum scales. 

In this note, we will shortly introduce a very recent proposal~\cite{Binosi:2016nme}, on the basis of our belief of a unique QCD running coupling, for a process-independent (PI) effective charge which is an analogue of the Gell-Mann-Low effective coupling in QED because it is completely determined by the gauge-boson propagator information. The evaluation of this PI effective charge will be also updated, by incorporating non-negligible effects on the gluon vacuum polarisation from ghost-gluon dynamics. Finally, the effective charge thus evaluated will be preliminarly applied to compute the proton Dirac electromagnetic form factor.

\section{PI effective charge}
\label{PI}

As in the case of QED, certainly owing to the uniqueness of the running coupling, a process-independent effective charge can be obtained simply by computing the photon vacuum polarisation. Indeed, in Abelian theories, there is no ghost sector (or it fully decouples from the theory dynamics) and one is provided with the Ward identity which guarantees that the electric-charge renormalisation constant is equivalent to that of the photon field; therefore deeply relating both the dressing of the vacuum polarisation and that of the interaction vertices. This paves the way for the effective charge definition. 

In QCD, a non-Abelian theory, one needs to deal with ghost fields. However, despite this intrinsic complexity, there is one approach to analysing QCD's Schwinger functions that preserves some of QED's Abelian simplicity: the combination of pinch technique (PT) \cite{Cornwall:1981zr,Cornwall:1989gv,Pilaftsis:1996fh,Binosi:2002ft,Binosi:2003rr,Binosi:2009qm} and background field method (BFM) \cite{Abbott:1980hw,Abbott:1981ke}.  The PT-BFM framework can be seen as a mean by which QCD can be `Abelianised" by the systematic rearranging of classes of diagrams and their sums in order to obtain modified Schwinger functions that satisfy linear (Abelian-like) STIs. Within this framework, In the gauge sector and in Landau gauge, it can be proved that all required features of the renormalisation group become captured by the gluon vacuum polarisation, and one can thus compute the QCD running coupling from the PT-BFM modified gluon dressing function. On top of this, the same result would be obtained, whichever the considered scattering process might be (gluon+gluon$\,\to\,$gluon+gluon, quark+quark$\,\to\,$quark+quark, etc.). It is worthwhile to highlight that this PT-BFM running coupling also capitalises on another particular feature of QCD which happens precisely in Landau gauge, for which the renormalisation constant of the gluon-ghost vertex is not only finite but unity \cite{Taylor:1971ff}. On the ground of the latter, the effective charge obtained from the PT-BFM gluon vacuum polarisation is directly connected with that deduced from the gluon-ghost vertex \cite{Aguilar:2009nf}, also called the ``Taylor coupling,'' $\alpha_{\rm T}$ \cite{Blossier:2011tf,Blossier:2012ef,Blossier:2013ioa}.

\subsection{The definition of the PI coupling}

The particular definition for this QCD PI effective charge works as follows. One should begin with~\cite{Binosi:2014aea,Binosi:2016xxu}: 
\begin{subequations}
\label{allhatd}
\begin{align}
\label{hatd}
\alpha(\zeta^2) D^{\rm PB}_{\mu\nu}(k;\zeta) & = \widehat{d}(k^2)\, T_{\mu\nu}(k)\,,\\
%
%
  \mathpzc{I}(k^2) :=k^2 \widehat{d}(k^2) &= \frac{\alpha_{\rm T}(k^2)}{[1-L(k^2;\zeta^2)F(k^2;\zeta^2)]^2}\,,
  \label{mathcalI}
\end{align}
\end{subequations}
with $T_{\mu\nu}(k)=\delta_{\mu\nu}-k_\mu k_\nu/k^2$; and where:
$\alpha(\zeta^2)=g^2(\zeta^2)/[4\pi]$, $\zeta$ is the renormalisation scale;
$D^{\rm PB}_{\mu\nu}$ is the PT-BFM gluon two-point function;
$\widehat{d}(k^2)$ is the RGI running-interaction discussed in Ref.\,\cite{Aguilar:2009nf};
$F$ is the dressing function for the ghost propagator;
and $L$ is a longitudinal piece of the gluon-ghost vacuum polarisation that vanishes at $k^2=0$.
In terms of these quantities, QCD's matter-sector gap equation can be written $(k=p-q)$
\begin{subequations}
\label{gendseN}
\begin{align}
S^{-1}(p) 
& = Z_2 \,(i\gamma\cdot p + m^{\rm bm}) + \Sigma(p)\,,\\
\Sigma(p)& =  Z_2\int^\Lambda_{dq}\!\!
4\pi \widehat{d}(k^2) \,T_{\mu\nu}(k)\gamma_\mu S(q) \hat\Gamma^a_\nu(q,p)\, ,
\end{align}
\end{subequations}
where the usual $Z_1 \Gamma^a_\nu$ has become  $Z_2 \hat\Gamma^a_\nu$, with the latter being a PT-BFM gluon-quark vertex that satisfies an Abelian-like Ward-Green-Takahashi identity \cite{Binosi:2009qm} and $Z_{1,2}$ are, respectively, the gluon-quark vertex and quark wave function renormalisation constants.

The RGI interaction, $\widehat{d}(k^2)$, given in Eqs.\,\eqref{allhatd} is a crucial piece for the QCD applications of the PT-BFM computational framework and one of the key ingredients for the PI coupling definition that we are here outlining. It has been recently computed on the basis of the most up-to-date lattice inputs for the gauge propagators~\cite{Binosi:2014aea,Binosi:2016xxu} and it made strongly explicit a remarkable feature of QCD; namely, the saturation of the interaction at infrared momenta:
\begin{equation}\label{eq:m0}
\widehat{d}(k^2=0) = \alpha(\zeta^2)/m_g^2(\zeta)=\alpha_0/m_0^2\,,
\end{equation}
where $\alpha_0:=\alpha(0) \approx 0.9\pi$, $m_0:=m_g(0) \approx m_p/2$, \emph{i.e}.\ the gluon sector of QCD is characterised by a nonperturbatively-generated infrared mass-scale \cite{Cucchieri:2007md,Cucchieri:2007rg,Aguilar:2008xm,Dudal:2008sp,Bogolubsky:2009dc,Aguilar:2012rz,Gao:2017uox}.  Keeping this in mind, we can define the following RGI function
\begin{align}
\label{calD}
\mathpzc{D}(k^2) & = \frac{\Delta_{\rm F}(k^2;\zeta)}{ m_0^2\Delta_{\rm F}(0;\zeta)} \,,
\end{align}
employing for $\Delta_{\rm F}$ a parametrisation of continuum- and/or lattice-QCD calculations of the canonical gluon two-point function built such that  
\begin{eqnarray}\label{eq:Dasym}
\frac 1 {\mathpzc{D}(k^2)} = \left\{ 
\begin{array}{lr}
m_0^2+\mathrm{O}(k^2 \ln{k^2}) & k^2 \gg m_0^2 \\
k^2+ \mathrm{O}(1) & k^2 \ll m_0^2
\end{array} 
\right. 
\end{eqnarray}
so that the nonperturbative IR behaviour is preserved and the UV anomalous dimension remains in $\widehat{d}(k^2)$. Using Eq.\,\eqref{calD},
\begin{equation}
\Sigma(p)  =   Z_2 \int^\Lambda_{dq}\!\!
4\pi \widehat{\alpha}_{\rm PI}(k^2) \mathpzc{D}_{\mu\nu}(k^2) \gamma_\mu S(q) \hat\Gamma^a_\nu(q,p)\,,
\end{equation}
where $\mathpzc{D}_{\mu\nu} =\mathpzc{D} T_{\mu\nu}$ and the dimensionless product
\begin{equation}
\widehat{\alpha}_{\rm PI}(k^2)  = \frac{\widehat{d}(k^2)} {\mathpzc{D}(k^2)}
\label{widehatalpha}
\end{equation}
is a RGI running-coupling (effective charge): by construction, $\widehat{\alpha}_{\rm PI}(k^2) = \mathpzc{I}(k^2)$ on $k^2\gg m_0^2$.

\subsection{The computation of the PI effective coupling}

\begin{figure}[t]
  \includegraphics[width=0.75\textwidth]{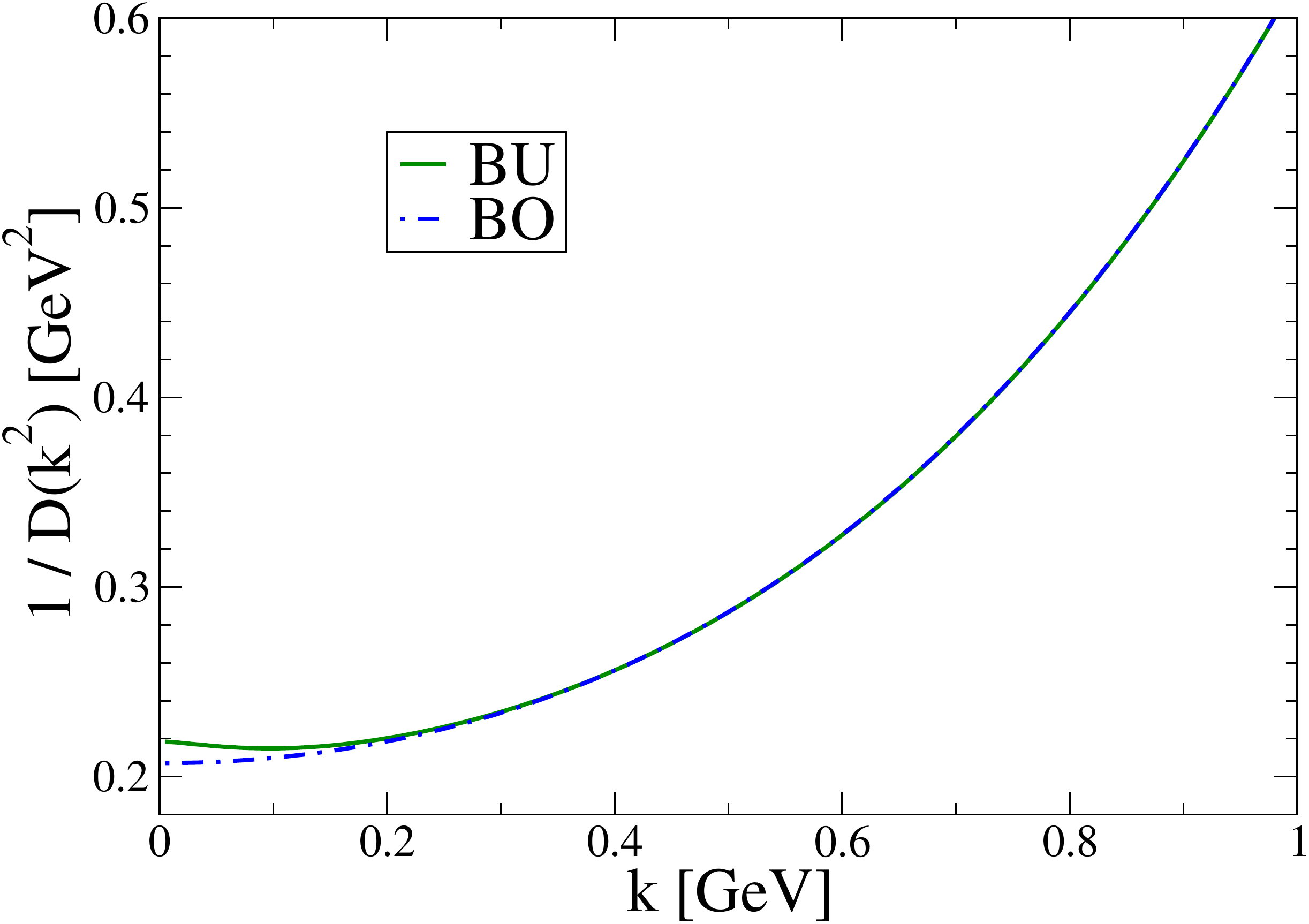}
\caption{The inverse of the RGI function $\mathpzc{D}(k^2)$, computed with a simple Pad\'e for $\Delta_F(k^2)$ as done in Ref.~\cite{Binosi:2016nme} (blue dot-dashed line) and by applying the interpolating function \eqref{eq:Deltacw} (green solid line), which amends the previous Pad\'e.} 
\label{fig:1}       
\end{figure}

\begin{figure}[t]
  \includegraphics[width=0.85\textwidth]{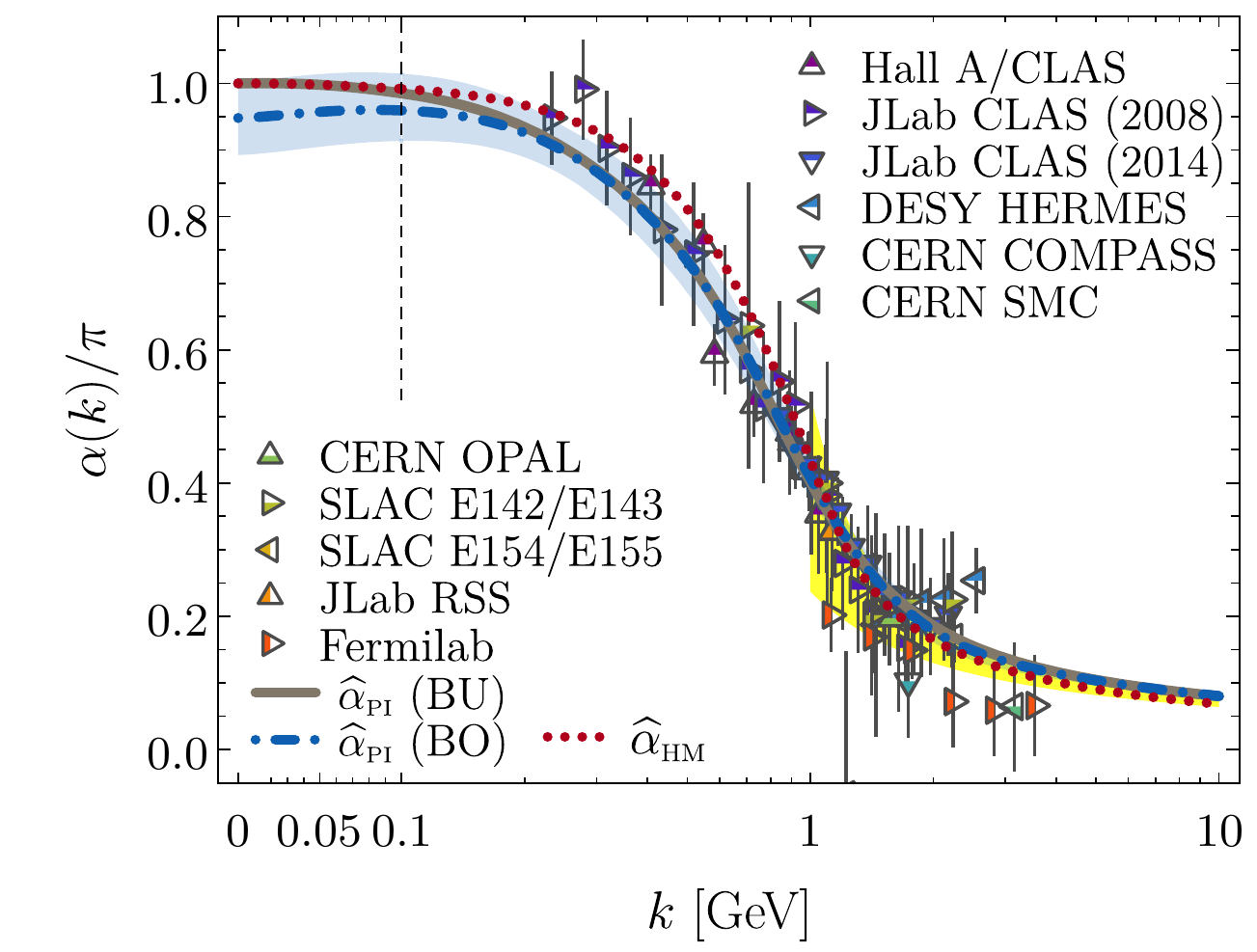}
\caption{Predicted process-in\-de\-pen\-dent RGI running-coupling $\widehat{\alpha}_{\rm PI}(k^2)$, Eq.\,\eqref{widehatalpha}, as it was obtained in Ref.~\cite{Binosi:2016nme} (dot-dashed blue curve) and as explained here (solid black curve): applying the amended interpolating function given by Eq.~\eqref{eq:Deltacw}, to account for lattice data. The shaded (blue) band bracketing the original curve combines a 95\% confidence-level window based on existing lattice-QCD results for the gluon two-point function with an error of 10\% in the continuum extraction of the RGI product $L F$ in Eqs.\,\eqref{allhatd}. 
World data on $\alpha_{g_1}$ (see Ref.~\cite{Binosi:2016nme} and references therein).
The shaded (yellow) band on $k>1\,$GeV represents $\alpha_{g_1}$ obtained from the Bjorken sum by using QCD evolution \cite{Gribov:1972,Altarelli:1977,Dokshitzer:1977} to extrapolate high-$k^2$ data into the depicted region, following Refs.\,\cite{Deur:2005cf,Deur:2008rf}; and, for additional context, the dotted (red) curve is the light-front holographic model of $\alpha_{g_1}$ canvassed in Ref.\,\cite{Deur:2016tte}.
%
} 
\label{fig:2}       
\end{figure}

The effective charge defined in Eq.\,\eqref{widehatalpha} results from a product of two known quantities; namely, $\widehat{d}(k^2)$  and the canonical gluon two-point function. Both have been extensively studied and tightly constrained using continuum and lattice methods and, very specially, known forms for both functions have been shown to provide a unified, quantitatively reliable explanation of numerous hadron physics observables \cite{Binosi:2014aea,Binosi:2016wcx}). The first quantity, $\widehat{d}(k^2)$, has been very recently computed \cite{Binosi:2016xxu}, for which contemporary results for the gluon propagator from lattice QCD are the only required input. In Ref.~\cite{Binosi:2016nme}, the same lattice results were also of help to obtain the RGI function $\mathpzc{D}(k^2)$, defined in Eq.\,\eqref{calD}, by the use of a  $[n,n+1]$ Pad\'e approximant\footnote{We used $n=1$ because $n\geq 2$ delivers no noticeable improvement while $n=0$ cannot account for modern lattice data} to simultaneously interpolate the IR behaviour of those lattice results~\cite{Binosi:2016xxu} and express the UV constraint on $\Delta_{\rm F}(k^2;\zeta)$ given by Eq.~\eqref{eq:Dasym}. The result for the RGI function $\mathpzc{D}(k^2)$ thus obtained appear displayed in Fig.~\ref{fig:1}, labelled as BO; and, combined with that of $\widehat{d}(k^2)$ from Ref.~\cite{Binosi:2016xxu} through Eq.~\eqref{widehatalpha}, they produce the results for the PI effective charge, also labelled as BO, in Fig.~\ref{fig:2}. This corresponds to the results published in Ref.~\cite{Binosi:2016nme}. 

%

However, as described in Ref.~\cite{Binosi:2016xxu}, the IR behavior of the gluon propagator is controlled by the appearence of the gluon mass-scale, $m_g^2(\zeta)$, \emph{viz}.\ at O$(k^2)$,
\begin{align}
\frac 1 {\Delta(k^2;\zeta^2)} &\underset{k^2\ll \zeta^2}{\approx} 
 m_g^2(\zeta^2) \left( 1 + \frac{ \mathpzc{l}_w(\zeta^2)}{m_g^2(\zeta^2)} \, k^2 \ln{\frac{k^2}{\zeta^2}} 
\, + \, \mathrm{O}(k^2) \right) \ ;
\label{gluonDelta}
\end{align}
where $\mathpzc{l}_w(\zeta^2)$ expresses the presence of massless-ghost loops in the gluon vacuum polarisation and has been shown to be  genuinely responsible for the zero-crossing of the three-gluon coupling at deeply low momenta~\cite{Aguilar:2013vaa,Tissier:2011ey,Pelaez:2013cpa,Athenodorou:2016oyh,Boucaud:2017obn}. Then, inasmuch as $\Delta_F$ preserves the nonpertubative IR behaviour of the gluon propagator,  it can be thereupon derived from Eqs.~\eqref{calD} and \eqref{gluonDelta} that 
\begin{align}
\frac 1 {\mathpzc{D}(k^2)} &\underset{k^2\ll \zeta^2}{\approx} 
 m_0^2 \left( 1 + \frac{ \mathpzc{l}_w(\zeta^2)}{m_g^2(\zeta^2)} \, k^2 \ln{\frac{k^2}{\Lambda_T^2}} 
\, + \, \mathrm{O}(k^2) \right) \ ,
\label{1overcalD}
\end{align}
where $m_0$ is the RGI mass-scale introduced in, and defined right after, Eq.~\eqref{eq:m0}, and the renormalisation scale, $\zeta$, has been appropriately traded for the fundamental $\Lambda_{\rm QCD}$ parameter defined in the Taylor scheme, $\Lambda_T$, for so to make explicit the RGI-nature for the expression inside the bracket of Eq.~\eqref{1overcalD}'s r.h.s. and, particularly, for $\mathpzc{l}_w/m_g^2$. In addition, also in Ref.~\cite{Binosi:2016xxu}, one is left with
\begin{align}
\widehat{d}(k^2) &\underset{k^2\ll \zeta^2}{\approx} 
 \widehat{d}(0) \left( 1 - \left( \frac{ \widehat{d}(0)}{8\pi} \, + \, \frac{ \mathpzc{l}_w(\zeta^2)}{m_g^2(\zeta^2)} \right) \, k^2 \ln{\frac{k^2}{\Lambda_T^2}} \, + \, \mathrm{O}(k^2) \right) \ .
\label{dhat}
\end{align}
Thus, Eqs.~\eqref{1overcalD} and \eqref{dhat} can be combined as dictated by Eq.~\eqref{widehatalpha} and so yield the following expression 
\begin{align}
\widehat{\alpha}_{\rm PI}(k^2) 
&\underset{k^2\ll \zeta^2}{\approx} 
 m_0^2 \, \widehat{d}(0) \left( 1 - \frac{ \widehat{d}(0)}{8\pi}  \, k^2 \ln{\frac{k^2}{\Lambda_T^2}} \, + \, \mathrm{O}(k^2) \right) \ ,
\label{alphaPI_IR}
\end{align}
describing the IR behaviour of the PI effective charge. The value for $\widehat{d}(0)$ is well established by the analysis of Ref.~\cite{Binosi:2016xxu} and appears to be 14.4 GeV$^{-2}$, while $m_0^2=1/\mathpzc{D}(0)$ keeps fully determined by the gluon propagator, here obtained by the interpolation of contemporary lattice data, used to build the RGI function $\mathpzc{D}(k^2)$ and have it obey the UV constraint given by Eq.~\eqref{eq:Dasym}. When using the Pad\'e approximant of Ref.~\cite{Binosi:2016nme}, one is left with $m_0=0.455$ GeV and the saturation point for the PI effective charge is found to be 
$\widehat{\alpha}_{\rm PI}(0)/\pi =0.949$, as can be seen in Fig.\ref{fig:2}. 

However, apart from the saturation point, the leading IR contribution for the running of $\widehat{\alpha}_{\rm PI}(k^2)$ is fully determined by $\widehat{d}(0)$. In particular, this leading correction, in competition with the $\mathrm{O}(k^2)$-term, makes the PI coupling to rise first, when the momentum increases from zero, reach then a maximum at a non-zero momentum and drop afterwards. This is a feature of the IR running for the PI effective charge that can be clearly noticed in the curve of Fig.~\ref{fig:2} which corresponds to using a simple Pad\'e for $\mathpzc{D}(k^2)$, as done in Ref.~\cite{Binosi:2016nme}. Though, the value we obtained for $\widehat{d}(0)$ appears not to be consistent with that curve. Indeed, the smaller is the coefficient in front of the logarithm the lower the momentum for which the maximum is reached; and our estimate for $\widehat{d}(0)$ comes to suggest that the maximum takes places at such a small value of momentum that it is scarcely noticeable and the coupling, in appearance, must behave monotonically. 

Therefore, the maximum in Fig.~\ref{fig:2} is nothing but an artifact that can be readily understood from Eq.~\eqref{gluonDelta}: the effect of the massless-ghost loops in the gluon vacuum polarisation translates into a negative singularity for the three-gluon vertex function and into the appearence of, again, a maximum at non-zero momentum~\cite{Aguilar:2013vaa,Athenodorou:2016oyh} for the gluon propagator. Whilst the simple Pad\'e used to interpolate the lattice data can potentially mimic this effect by a fit, for so to happen one would need to have enough data available at deeply low momenta, around the location of the maximum. And this is not the case for the simulations, including four dynamical quark flavours, exploited in Ref.~\cite{Binosi:2016wcx}. This issue can be simply addressed by using the following interpolation function
\begin{equation}
\label{eq:Deltacw}
\frac 1 {\Delta_F(k^2,\zeta^2)} \ = \ \frac{1+b_l \ k^2 \ln{\frac{k^2}{k^2+p_0^2}} + b_1 k^2 + b_2 k^4}{a_0 + a_1 k^2} \ , 
\end{equation}
which amends the Pad\'e by imposing the known asymptotic behaviour shown by Eq.~\eqref{gluonDelta}, and where $b_1$, $b_2$, $a_0$ and $a_1$ are free parameters to fit the lattice data, while $b_l=\mathpzc{l}_w(\zeta^ 2)/m^ 2_g(\zeta^ 2)= 2.5$ GeV$^{-2}$ is fixed with by the analysis of the three-gluon vertex function in Refs.~\cite{Athenodorou:2016oyh,Boucaud:2017obn} and $p_0=0.15$ GeV as it appears suggested by the position of the zero crossing identified for this vertex function~\cite{Boucaud:2017obn,Blum:2014gna,Eichmann:2014xya,Cyrol:2016tym}, which is related to the position of the maximum in the gluon propagator (both phenomena are intimately connected). Then, so proceeding for the interpolation of the lattice data and applying Eq.~\eqref{calD}, we obtain the curve labelled as BU in Fig.~\ref{fig:1} for the RGI function $\mathpzc{D}(k^2)$, which only deviates from that of Ref.~\cite{Binosi:2016nme} at very low momenta. This small IR effect comes however to enhance slightly the RGI mass scale, $m_0=0.468$ GeV, and this results in $\widehat{\alpha}_{\rm PI}(0)/\pi = m_0^2 \, \widehat{d}(0) /\pi = 1.002$ for the saturation point of the PI effective coupling. The full result appear displayed in Fig.~\ref{fig:2} through a curve labelled as BU and it clearly shows an apparent monotonically decreasing IR running, as expected. 

\subsection{Ansatz for the PI effective coupling}

Capitalising on the well-known asymptotics of $\widehat{\alpha}_{\rm PI}(k^2)$ in both the IR (see Eq.~\eqref{alphaPI_IR}) and UV domains\footnote{Eq.~\eqref{mathcalI} establishes a direct connection with the well-known Taylor coupling}, the numerical results for the PI effective coupling displayed in Fig.~\ref{fig:2} can be very accurately described with the following ansatz: 
\begin{equation}\label{eq:ansf}
\frac {\widehat{\alpha}_{\rm PI}(k^2)}{\pi} \ = \ 
\frac{\displaystyle 1 - \frac{\widehat{d}(0)}{8\pi} k^2 \ln\frac{k^2}{\Lambda_T^2} +  d_1 k^2 + d_2 k^4}
{\displaystyle 1 + b_1 k^2 +  c_0 d_2  \ k^4 \ln\frac{k^2}{\Lambda_T^2}} \ ; 
\end{equation}
which is molded to satisfy Eq.~\eqref{alphaPI_IR} at $k^2 \gg \Lambda^2_T$ and to behave as $1/[c_0 \ln{(k^2/\Lambda^2_T)}]$, where $c_0=(11-2 N_f/3)/(4\pi)$ and $\Lambda_T=500$ MeV, at  $k^2 \ll \Lambda^2_T$. We know from numerical computation that $\widehat{d}(0)=14.4$ GeV$^{-2}$ and have already used $\widehat{\alpha}_{\rm PI}(0)=\pi$, as a very good approximation to the zero-momentum saturation value that we computed above. The fit of Eq.~\eqref{eq:ansf} to the numerical results displayed in Fig.~\ref{fig:2} by the curve labelled as BU produces a pointwise identical curve when $d_1$=3.56 GeV$^{-2}$, $d_2$=2.85 GeV$^{-4}$ and $b_1$=7.25 GeV$^{-2}$.

\section{Comparison of effective charges}

There are other different approaches to determining ``effective charges'' in QCD, as that introduced in Ref.\,\cite{Grunberg:1982fw}. 
This last is based on a genuinely process-dependent procedure, where the effective running coupling is defined to be completely fixed by the leading-order term in the perturbative expansion of a given observable in terms of the canonical running coupling. Naturally, effective charges from different observables can in principle be algebraically connected to each other via an expansion of one coupling in terms of the other. However, any such expansion can only be defined \emph{a posteriori}, \emph{i.e}.\ after both effective charges are independently constructed, and contains infinitely many terms \cite{Deur:2016tte}.

One such process-dependent effective charge is $\alpha_{g_1}(k^2)$, which is defined via the Bjorken sum rule \cite{Bjorken:1966jh,Bjorken:1969mm}:
\begin{align}
\int_0^1 \! dx  \left[g_1^p(x,k^2) - g_1^n(x,k^2)\right] = \frac{g_A}{6}  \left[ 1 - \tfrac{1}{\pi} \alpha_{g_1}(k^2) \right]\,,
\end{align}
where $g_1^{p,n}$ are the spin-dependent proton and neutron structure functions, whose extraction requires measurements using polarised targets, and $g_A$ is the nucleon flavour-singlet axial-charge \cite{Aidala:2012mv}. This particular definition is endowed with merits that are outlined in Ref.\,\cite{Deur:2016tte} and that make interesting the comparison with our PI effective coupling. This comparison can be seen in Fig.~\ref{fig:2}, where the world's data on the process-dependent effective charge $\alpha_{g_1}(k^2)$ are depicted. 
It should be highlighted that, expanded asymptotically in terms of the widely-used $\overline{\rm MS}$ running coupling~\cite{Olive:2016xmw}, both definitions agree very well in the UV domain:
\begin{subequations}
\label{AgreeCouplings}
 \begin{align}
 \label{ag1MSbar}
 \alpha_{g_1}(k^2) & = \alpha_{\overline{\rm MS}}(k^2) ( 1 + 1.14 \, \alpha_{\overline{\rm MS}}(k^2) + \ldots ) \,,\\
 \widehat{\alpha}_{\rm PI}(k^2) & = \alpha_{\overline{\rm MS}}(k^2) ( 1 + 1.09 \, \alpha_{\overline{\rm MS}}(k^2) + \ldots ) \,,
 \end{align}
\end{subequations}
where Eq.\,\eqref{ag1MSbar} may be built from, \emph{e.g}.\ Refs.\,\cite{Kataev:1994gd,Baikov:2010je}. Furthermore, apart from this perturbative near coincidence, there is also near precise agreement with data on the IR domain, $k^2 \lesssim m_0^2$, and complete accord on $k^2 \geq m_0^2$. This is a significant result supporting our proposal based on the gauge-sector information within the PT-BFM framework, in analogy with the QED Gell-Mann-Low effective coupling. It is highly worthwhile to emphasize that any agreement on $k^2\in[0.01,1]\,$GeV$^2$ is non-trivial because ghost-gluon interactions produce as much as 40\% of $\widehat{\alpha}_{\rm PI}(k^2)$ on this domain. 
In Fig.~\ref{fig:2}, a comparison with the light-front holographic model of $\alpha_{g_1}$ canvassed in Ref.\,\cite{Deur:2016tte} is also made. 

\section{Proton form factor}

\begin{figure}
  \includegraphics[width=0.95\textwidth]{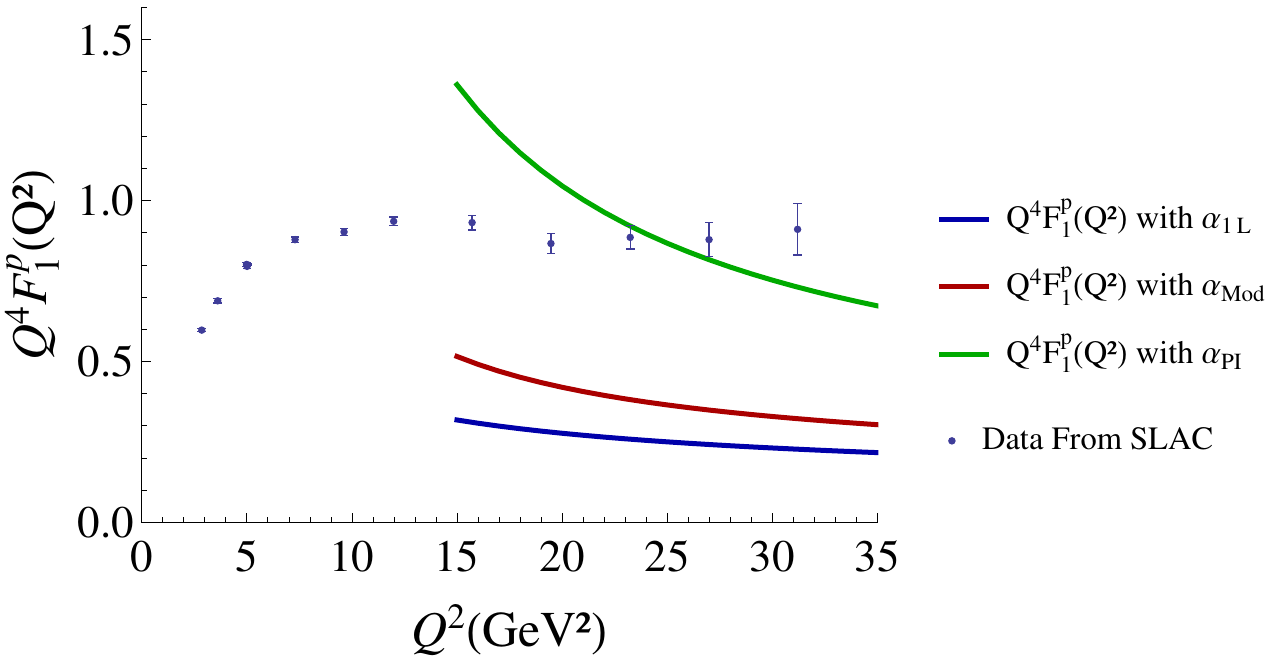}
\caption{Proton electromagnetic form factor computed by Eqs.~\eqref{eq:FormFactorGeneral} and \eqref{eq:THExpansion} with the PDAs derived in Ref.~\cite{Mezrag:2017znp}, a one-loop perturbative running coupling (solid blue curve), the PI effective coupling here introduced, expressed by the ansatz in \eqref{eq:ansf} (solid green curve) and the same ansatz but replacing $\Lambda_T$ by $\Lambda_{\overline{\rm MS}}$=0.234 GeV (red solid curve).  
} 
\label{fig:3}       
\end{figure}

As a very preliminary example of a phenomenological application for our PI effective coupling, aiming at nothing but sketching the rough impact it might have, we consider here the nucleon electromagnetic form factor $F_1$ that, in the hard scattering regime, can be written as: 
\begin{eqnarray}
  \label{eq:FormFactorGeneral}
  && Q^4 F_1(Q^2) = \frac{8\pi^2}{27} \int [\textrm{d}x] \int [\textrm{d}y] \nonumber \\ 
  &&\rule[0cm]{1cm}{0cm} \times \, f_n(\zeta_x^2) \varphi(x_i,\zeta_x) T_H \left( x_i,y_i, Q^2, \zeta_x^2,\zeta_y^2,\mu^2 \right) f_n(\zeta_y^2)\varphi(y_i,\zeta_y) ,
\end{eqnarray}
where $\varphi$ is the nucleon parton distribution amplitude (PDA) evaluated at the scale $\zeta$, $f_n$ is the nucleon normalisation and $T_H$ the hard scattering kernel that one can compute using perturbation theory. The $\zeta$'s correspond to the ``left'' and ``right'' factorisation scales handling with IR singularities, while $\mu$ is the renormalisation scale dealing with the UV singularities. $Q^2$ is the photon virtuality. $T_H$ can be then calculated through perturbation theory, and one is thus left with
\begin{eqnarray}
  \label{eq:THExpansion}
  T_H(x_i,y_i,Q^2,\zeta_x,\zeta_y,\mu) &=& T^0(x_i,y_i,Q^2,\mu) \nonumber \\
  &\times& \left(1 + \alpha_S(\mu)T^1(x_i,y_i,Q^2,\zeta_x,\zeta_y,\mu) + \dots \,\right).
\end{eqnarray}
The results for $T^0$ are given in different papers \cite{Lepage:1980fj,Chernyak:1984bm,Ji:1986uh}, with different ways to handle the scales. For the illustrative purposes we pursue, we will content ourselves with using the prescription~\cite{Chernyak:1984bm} of taking the coupling at the typical momentum of the gluon and freezing the scale of the PDAs to 1 GeV. We thus apply the nucleon PDAs for the proton recently obtained in Ref.~\cite{Mezrag:2017znp} (see also C.D. Roberts's contribution to these proceedings) and compute the electromagnetic proton form factor by considering a one-loop perturbative running coupling and the ansatz given by Eq.~\eqref{eq:ansf} for the PI effective coupling introduced here. The results, compared to SLAC data, can be seen in Fig.~\ref{fig:3}. 

\section{Conclusions}

We have defined a new type of effective charge in QCD, which is an analogue of the QED Gell-Mann-Low effective coupling, being completely determined by the gauge-field two-point Green's function, capitalising on the PT-BFM computational framework which preserves some of QED's simplicities. We have thus calculated a running-coupling for QCD which has the merit of being process-independent and parameter-free, being obtained by combining the self-consistent solution of a set of Dyson-Schwinger equations with inputs from lattice-QCD; and it smoothly unifies the nonperturbative and perturbative domains of the strong-interaction theory. 
%

Compared to the process-dependent effective charge, $\alpha_{g_1}$,  defined via the Bjorken sum rule, our process-independent coupling is almost pointwise identical at infrared momenta and, even more, is very nearly satisfying $\widehat{\alpha}_{\rm PI}(0)=\pi$, as required by the Bjorken sum rule, after properly accounting for a non-negligible contribution from the ghost-gluon dynamics in the gluon vacuum polarisation which, basically, impacts slightly on the gluon-sector nonpertur\-batively-generated IR mass scale and on its running at deeply low momenta. Our predicted running coupling is free of Landau pole and possesses an inflection point at $\surd k^2=0.7\,$GeV, at which the growth of the coupling when momenta decreases from UV to IR starts to slow, that coupling turns away from the Landau pole and finally saturates: $\widehat\alpha_{\rm PI}(k^2=0) \approx \pi$.  

Finally, we have also shown a very preliminary calculation for the proton electromagnetic form factor in the hard scattering regime, plugging our effective charge into the hard scattering kernel, aiming only at a rough estimate of its impact on the results; and finding it to be sizeable and in the right direction, according to the available empirical results.

\begin{acknowledgements}
We are grateful for comments from S.\,J.\,Brodsky, L.\,Chang, A.\,Deur and S.-X.\,Qin.
%
This study was conceived and initiated during the $3^{rd}$ Workshop on Non-perturbative QCD, University of Seville, Spain, 17-21 October 2016.
This research was supported by:
Spanish MINECO, under grants FPA2014-53631-C-1-P, FPA2014-53631-C-2-P, FPA2017-86380-P and SEV-2014-0398;
Generalitat Valenciana  under grant Prometeo~II/2014/066;
and
U.S.\ Department of Energy, Office of Science, Office of Nuclear Physics, contract no.~DE-AC02-06CH11357.
\end{acknowledgements}


\begin{thebibliography}{10}
\providecommand{\url}[1]{{#1}}
\providecommand{\urlprefix}{URL }
\expandafter\ifx\csname urlstyle\endcsname\relax
  \providecommand{\doi}[1]{DOI \discretionary{}{}{}#1}\else
  \providecommand{\doi}{DOI \discretionary{}{}{}\begingroup
  \urlstyle{rm}\Url}\fi

\bibitem{Binosi:2016nme}
D.~Binosi, C.~Mezrag, J.~Papavassiliou, C.D. Roberts, J.~Rodriguez-Quintero,
  Phys. Rev. \textbf{D96}(5), 054026 (2017).
\newblock \doi{10.1103/PhysRevD.96.054026}

\bibitem{Ward:1950xp}
J.C. Ward, Phys. Rev. \textbf{78}, 182 (1950).
\newblock \doi{10.1103/PhysRev.78.182}

\bibitem{Marciano:1979wa}
W.J. Marciano, H.~Pagels, Nature \textbf{279}, 479 (1979).
\newblock \doi{10.1038/279479a0}

\bibitem{Becchi:1975nq}
C.~Becchi, A.~Rouet, R.~Stora, Annals Phys. \textbf{98}, 287 (1976).
\newblock \doi{10.1016/0003-4916(76)90156-1}

\bibitem{Tyutin:1975qk}
I.V. Tyutin,   (1975)

\bibitem{Taylor:1971ff}
J.~Taylor, Nucl.Phys. \textbf{B33}, 436 (1971).
\newblock \doi{10.1016/0550-3213(71)90297-5}

\bibitem{Slavnov:1972fg}
A.~Slavnov, Theor.Math.Phys. \textbf{10}, 99 (1972).
\newblock \doi{10.1007/BF01090719, 10.1007/BF01090719}

\bibitem{Cucchieri:2007md}
A.~Cucchieri, T.~Mendes, PoS \textbf{LAT2007}, 297 (2007)

\bibitem{Cucchieri:2007rg}
A.~Cucchieri, T.~Mendes, Phys.Rev.Lett. \textbf{100}, 241601 (2008).
\newblock \doi{10.1103/PhysRevLett.100.241601}

\bibitem{Aguilar:2008xm}
A.C. Aguilar, D.~Binosi, J.~Papavassiliou, Phys. Rev. \textbf{D78}, 025010
  (2008).
\newblock \doi{10.1103/PhysRevD.78.025010}

\bibitem{Boucaud:2008ky}
P.~Boucaud, J.~Leroy, A.~Le~Yaouanc, J.~Micheli, O.~P\`ene, et~al., JHEP
  \textbf{0806}, 099 (2008).
\newblock \doi{10.1088/1126-6708/2008/06/099}

\bibitem{Dudal:2008sp}
D.~Dudal, J.A. Gracey, S.P. Sorella, N.~Vandersickel, H.~Verschelde, Phys. Rev.
  \textbf{D78}, 065047 (2008).
\newblock \doi{10.1103/PhysRevD.78.065047}

\bibitem{Bogolubsky:2009dc}
I.~Bogolubsky, E.~Ilgenfritz, M.~Muller-Preussker, A.~Sternbeck, Phys.Lett.
  \textbf{B676}, 69 (2009).
\newblock \doi{10.1016/j.physletb.2009.04.076}

\bibitem{Aguilar:2012rz}
A.C. Aguilar, D.~Binosi, J.~Papavassiliou, Phys. Rev. \textbf{D86}, 014032
  (2012).
\newblock \doi{10.1103/PhysRevD.86.014032}

\bibitem{Bhagwat:2003vw}
M.~Bhagwat, M.~Pichowsky, C.~Roberts, P.~Tandy, Phys.Rev. \textbf{C68}, 015203
  (2003).
\newblock \doi{10.1103/PhysRevC.68.015203}

\bibitem{Bowman:2005vx}
P.O. Bowman, U.M. Heller, D.B. Leinweber, M.B. Parappilly, A.G. Williams,
  et~al., Phys.Rev. \textbf{D71}, 054507 (2005).
\newblock \doi{10.1103/PhysRevD.71.054507}

\bibitem{Bhagwat:2006tu}
M.~Bhagwat, P.~Tandy, AIP Conf.Proc. \textbf{842}, 225 (2006).
\newblock \doi{10.1063/1.2220232}

\bibitem{Cornwall:1981zr}
J.M. Cornwall, Phys.Rev. \textbf{D26}, 1453 (1982).
\newblock \doi{10.1103/PhysRevD.26.1453}

\bibitem{Cornwall:1989gv}
J.M. Cornwall, J.~Papavassiliou, Phys. Rev. \textbf{D40}, 3474 (1989).
\newblock \doi{10.1103/PhysRevD.40.3474}

\bibitem{Pilaftsis:1996fh}
A.~Pilaftsis, Nucl. Phys. \textbf{B487}, 467 (1997).
\newblock \doi{10.1016/S0550-3213(96)00686-4}

\bibitem{Binosi:2002ft}
D.~Binosi, J.~Papavassiliou, Phys.Rev. \textbf{D66}, 111901 (2002).
\newblock \doi{10.1103/PhysRevD.66.111901}

\bibitem{Binosi:2003rr}
D.~Binosi, J.~Papavassiliou, J. Phys. \textbf{G30}, 203 (2004).
\newblock \doi{10.1088/0954-3899/30/2/017}

\bibitem{Binosi:2009qm}
D.~Binosi, J.~Papavassiliou, Phys.Rept. \textbf{479}, 1 (2009).
\newblock \doi{10.1016/j.physrep.2009.05.001}.
\newblock 245 pages, 92 figures

\bibitem{Abbott:1980hw}
L.F. Abbott, Nucl. Phys. \textbf{B185}, 189 (1981).
\newblock \doi{10.1016/0550-3213(81)90371-0}

\bibitem{Abbott:1981ke}
L.F. Abbott, Acta Phys. Polon. \textbf{B13}, 33 (1982)

\bibitem{Aguilar:2009nf}
A.C. Aguilar, D.~Binosi, J.~Papavassiliou, J.~Rodriguez-Quintero, Phys. Rev.
  \textbf{D80}, 085018 (2009).
\newblock \doi{10.1103/PhysRevD.80.085018}

\bibitem{Blossier:2011tf}
B.~Blossier, P.~Boucaud, M.~Brinet, F.~De~Soto, X.~Du, et~al., Phys.Rev.
  \textbf{D85}, 034503 (2012).
\newblock \doi{10.1103/PhysRevD.85.034503}

\bibitem{Blossier:2012ef}
B.~Blossier, P.~Boucaud, M.~Brinet, F.~De~Soto, X.~Du, V.~Morenas, O.~Pene,
  K.~Petrov, J.~Rodriguez-Quintero, Phys. Rev. Lett. \textbf{108}, 262002
  (2012).
\newblock \doi{10.1103/PhysRevLett.108.262002}

\bibitem{Blossier:2013ioa}
B.~Blossier, et~al., Phys.Rev. \textbf{D89}, 014507 (2014).
\newblock \doi{10.1103/PhysRevD.89.014507}

\bibitem{Binosi:2014aea}
D.~Binosi, L.~Chang, J.~Papavassiliou, C.D. Roberts, Phys. Lett. \textbf{B742},
  183 (2015).
\newblock \doi{10.1016/j.physletb.2015.01.031}

\bibitem{Binosi:2016xxu}
D.~Binosi, C.D. Roberts, J.~Rodriguez-Quintero, Phys. Rev. \textbf{D95}(11),
  114009 (2017).
\newblock \doi{10.1103/PhysRevD.95.114009}

\bibitem{Gao:2017uox}
F.~Gao, S.X. Qin, C.D. Roberts, J.~Rodriguez-Quintero,   (2017)

\bibitem{Binosi:2016wcx}
D.~Binosi, L.~Chang, J.~Papavassiliou, S.X. Qin, C.D. Roberts, Phys. Rev.
  \textbf{D95}(3), 031501 (2017).
\newblock \doi{10.1103/PhysRevD.95.031501}

\bibitem{Gribov:1972}
V.N. Gribov, L.N. Lipatov, Sov. J. Nucl. Phys. \textbf{15}, 438 (1972).
\newblock [Yad. Fiz.15,781(1972)]

\bibitem{Altarelli:1977}
G.~Altarelli, G.~Parisi, Nucl. Phys. \textbf{B126}, 298 (1977).
\newblock \doi{10.1016/0550-3213(77)90384-4}

\bibitem{Dokshitzer:1977}
Y.L. Dokshitzer, Sov. Phys. JETP \textbf{46}, 641 (1977).
\newblock [Zh. Eksp. Teor. Fiz.73,1216(1977)]

\bibitem{Deur:2005cf}
A.~Deur, V.~Burkert, J.P. Chen, W.~Korsch, Phys. Lett. \textbf{B650}, 244
  (2007).
\newblock \doi{10.1016/j.physletb.2007.05.015}

\bibitem{Deur:2008rf}
A.~Deur, V.~Burkert, J.P. Chen, W.~Korsch, Phys. Lett. \textbf{B665}, 349
  (2008).
\newblock \doi{10.1016/j.physletb.2008.06.049}

\bibitem{Deur:2016tte}
A.~Deur, S.J. Brodsky, G.F. de~Teramond, Prog. Part. Nucl. Phys. \textbf{90}, 1
  (2016).
\newblock \doi{10.1016/j.ppnp.2016.04.003}

\bibitem{Aguilar:2013vaa}
A.C. Aguilar, D.~Binosi, D.~Ibañez, J.~Papavassiliou, Phys. Rev.
  \textbf{D89}(8), 085008 (2014).
\newblock \doi{10.1103/PhysRevD.89.085008}

\bibitem{Tissier:2011ey}
M.~Tissier, N.~Wschebor,   (2011)

\bibitem{Pelaez:2013cpa}
M.~Pelaez, M.~Tissier, N.~Wschebor, Phys. Rev. \textbf{D88}, 125003 (2013).
\newblock \doi{10.1103/PhysRevD.88.125003}

\bibitem{Athenodorou:2016oyh}
A.~Athenodorou, D.~Binosi, P.~Boucaud, F.~De~Soto, J.~Papavassiliou,
  J.~Rodriguez-Quintero, S.~Zafeiropoulos, Phys. Lett. \textbf{B761}, 444
  (2016).
\newblock \doi{10.1016/j.physletb.2016.08.065}

\bibitem{Boucaud:2017obn}
P.~Boucaud, F.~De~Soto, J.~Rodríguez-Quintero, S.~Zafeiropoulos, Phys. Rev.
  \textbf{D95}(11), 114503 (2017).
\newblock \doi{10.1103/PhysRevD.95.114503}

\bibitem{Blum:2014gna}
A.~Blum, M.Q. Huber, M.~Mitter, L.~von Smekal, Phys. Rev. \textbf{D89}, 061703
  (2014).
\newblock \doi{10.1103/PhysRevD.89.061703}

\bibitem{Eichmann:2014xya}
G.~Eichmann, R.~Williams, R.~Alkofer, M.~Vujinovic, Phys. Rev.
  \textbf{D89}(10), 105014 (2014).
\newblock \doi{10.1103/PhysRevD.89.105014}

\bibitem{Cyrol:2016tym}
A.K. Cyrol, L.~Fister, M.~Mitter, J.M. Pawlowski, N.~Strodthoff, Phys. Rev.
  \textbf{D94}(5), 054005 (2016).
\newblock \doi{10.1103/PhysRevD.94.054005}

\bibitem{Grunberg:1982fw}
G.~Grunberg, Phys. Rev. \textbf{D29}, 2315 (1984).
\newblock \doi{10.1103/PhysRevD.29.2315}

\bibitem{Bjorken:1966jh}
J.D. Bjorken, Phys. Rev. \textbf{148}, 1467 (1966).
\newblock \doi{10.1103/PhysRev.148.1467}

\bibitem{Bjorken:1969mm}
J.D. Bjorken, Phys. Rev. \textbf{D1}, 1376 (1970).
\newblock \doi{10.1103/PhysRevD.1.1376}

\bibitem{Aidala:2012mv}
C.A. Aidala, S.D. Bass, D.~Hasch, G.K. Mallot, Rev. Mod. Phys. \textbf{85}, 655
  (2013).
\newblock \doi{10.1103/RevModPhys.85.655}

\bibitem{Olive:2016xmw}
C.~Patrignani, et~al., Chin. Phys. \textbf{C40}(10), 100001 (2016).
\newblock \doi{10.1088/1674-1137/40/10/100001}

\bibitem{Kataev:1994gd}
A.L. Kataev, Phys. Rev. \textbf{D50}, R5469 (1994).
\newblock \doi{10.1103/PhysRevD.50.R5469}

\bibitem{Baikov:2010je}
P.A. Baikov, K.G. Chetyrkin, J.H. Kuhn, Phys. Rev. Lett. \textbf{104}, 132004
  (2010).
\newblock \doi{10.1103/PhysRevLett.104.132004}

\bibitem{Lepage:1980fj}
G.P. Lepage, S.J. Brodsky, Phys. Rev. \textbf{D22}, 2157 (1980).
\newblock \doi{10.1103/PhysRevD.22.2157}

\bibitem{Chernyak:1984bm}
V.L. Chernyak, I.R. Zhitnitsky, Nucl. Phys. \textbf{B246}, 52 (1984).
\newblock \doi{10.1016/0550-3213(84)90114-7}

\bibitem{Ji:1986uh}
C.R. Ji, A.F. Sill, R.M. Lombard, Phys. Rev. \textbf{D36}, 165 (1987).
\newblock \doi{10.1103/PhysRevD.36.165}

\bibitem{Mezrag:2017znp}
C.~Mezrag, J.~Segovia, L.~Chang, C.D. Roberts,   (2017)

\end{thebibliography}

%
%

\end{document}